\documentclass[sigconf,nonacm]{acmart}

\setcopyright{none}
\renewcommand\footnotetextcopyrightpermission[1]{}

\AtBeginDocument{%
}

\usepackage{enumitem}
\usepackage{booktabs}
\usepackage{multirow}
\usepackage{colortbl}
\usepackage[table]{xcolor}
\usepackage{makecell}

\begin{document}

\title{When Context Bites: Detecting RAG Poisoning via Document-Level Attention Collapse}

\author{Yingtao Ren}
\authornote{Corresponding author}
\email{yingtao.ren@student.uts.edu.au}
\orcid{0000-0003-2035-4845}
\affiliation{%
  \institution{University of Technology Sydney}
  \city{Sydney}
  \country{Australia}
}

\author{Ziyi Zhao}
\email{ziyi.zhao-2@student.uts.edu.au}
\orcid{0000-0003-3537-8065}
\affiliation{%
  \institution{University of Technology Sydney}
  \city{Sydney}
  \country{Australia}
}

\author{Yiwei Fu}
\email{fuyw@stu.pku.edu.cn}
\orcid{0009-0000-1637-6764}
\affiliation{%
  \institution{Peking University}
  \city{Beijing}
  \country{China}
}

\author{Xiao Luo}
\email{xiao.luo@wisc.edu}
\orcid{0000-0002-7987-3714}
\affiliation{%
  \institution{University of Wisconsin--Madison}
  \city{Madison}
  \country{USA}
}

\author{Yu-Cheng Chang}
\email{fred.chang@uts.edu.au}
\orcid{0000-0001-9244-0318}
\affiliation{%
  \institution{University of Technology Sydney}
  \city{Sydney}
  \country{Australia}
}

\author{Chin-Teng Lin}
\authornotemark[1]
\email{chin-teng.lin@uts.edu.au}
\orcid{0000-0001-8371-8197}
\affiliation{%
  \institution{University of Technology Sydney}
  \city{Sydney}
  \country{Australia}
}

\renewcommand{\shortauthors}{Yingtao Ren et al.}

\begin{abstract}
Retrieval-augmented generation (RAG) is indispensable for enhancing large language models. However, RAGs are increasingly susceptible to poisoning attacks, in which adversarial documents are injected to manipulate generator outputs. Previous methods rely on output-side signals such as perplexity and consistency checks to detect such attacks. Nevertheless, our analysis reveals that deliberate attacks often induce false confidence, where poisoned outputs exhibit even lower perplexity than benign ones, rendering uncertainty-based detection ineffective. To address this challenge, we explore the internal dynamics of the generator and identify a distinctive signature termed \textit{Attention Collapse}. Unlike the dispersed attention in benign generations, attacked generations exhibit a decrease in entropy as attention concentrates on poisoned documents. Building on these findings, we propose \texttt{D-SCAN} (Document-level Signal Collapse Analysis), a lightweight detection framework that monitors attention dynamics to identify attacked generations. Extensive experiments on multiple attack benchmarks demonstrate the effectiveness of our method. Moreover, D-SCAN can detect attacks even when they fail to alter the final answer. Code is available at \url{https://github.com/yingtaoren/D-Scan.git}.

\end{abstract}


\maketitle

\section{Introduction}

Retrieval-Augmented Generation effectively bridges knowledge gaps in large language models (LLMs) through external retrieval~\cite{reasoningRAGL,llmagentL,fan2024ragsurvey}. Retrieved knowledge helps LLMs access the latest information that is not included in internal parameters, but it remains prone to injecting poisoning attacks~\citep{iclr_irrelevant,rbft,add_prompt_inject}. In such scenarios, attackers inject adversarial documents into the retrieved database to manipulate the generator to produce attacker-specified harmful responses~\citep{inference_time,care_paper}. This vulnerability poses unacceptable risks in high-stakes QA systems, such as legal and medical domains~\citep{LLM4scienceL,fu2025mark,hallunicationllm}. 

In practical deployments, detecting poisoning attacks and hallucinations primarily relies on post-hoc inspection of the final generated text, which is neither efficient nor accurate. Therefore, recent RAG-based detection works have turned to check implicit output-side generation signals~\citep{detect_survey}, employing uncertainty or consistency metrics to assess answers~\citep{reppl}, or training black-box classifiers to detect anomalies~\citep{inside,eigentrack}. However, these methods suffer from two fundamental limitations. First, they neglect the granular nature of RAG context, in which information is sparsely distributed across discrete retrieved chunks. Second, such approaches lack interpretability, offering only binary decisions without exploring the internal mechanisms of generations process.

\begin{figure*}[ht]
    \centering
    \setlength{\abovecaptionskip}{0.1pt}
    \includegraphics[width=\textwidth]{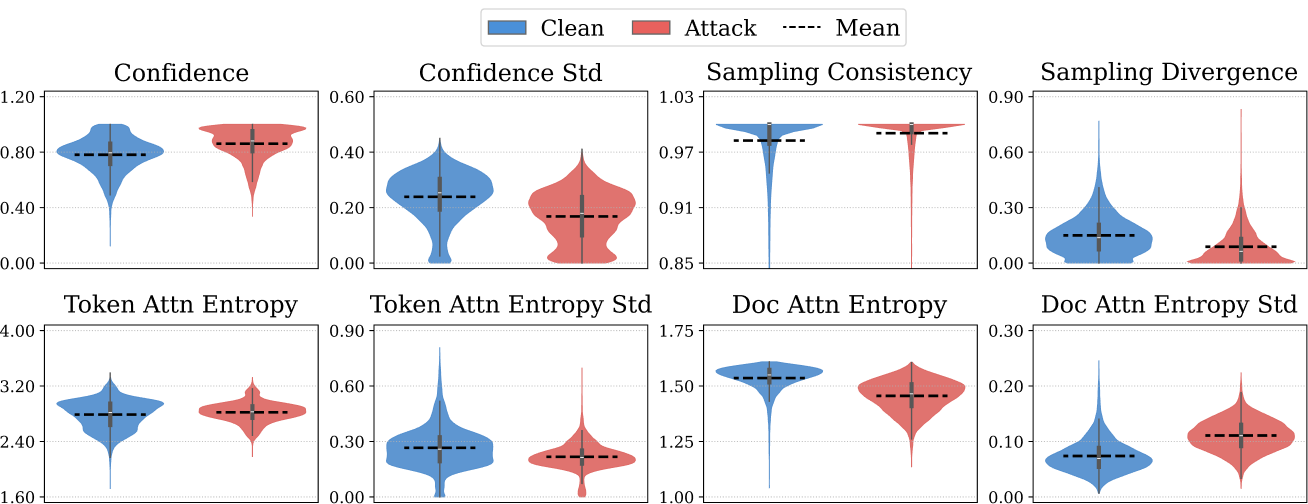}
    \vspace{-4pt}
    \caption{
    Comparison of internal dynamics of LLM between clean and poisoned generation. Reveals the blind confidence and attention collapse phenomenon. The attention collapse is more pronounced at the document level.
    }
    \label{fig:violin}
\end{figure*}

To address these limitations, we leverage mechanistic interpretability to investigate the internal generation dynamics~\cite{inner_state_acl} of the LLM generator under adversarial poisoning attacks. Our empirical analysis reveals a counterintuitive phenomenon regarding output-side signals. Adversarial contexts are specifically optimized to maximize the likelihood of generation, thereby inducing a state of false confidence. Consequently, poisoned responses often exhibit higher average token probabilities than clean responses. This overconfidence renders traditional uncertainty-based metrics and consistency checks ineffective, as the model consistently and confidently generates the attacker's target outputs. Furthermore, we observe that during poisoning attacks, the model's attention is hijacked by the carefully crafted poisoning content. We term this \textit{Attention Collapse}. Unlike benign generations that allocate dispersed attention to each relevant document, the attacked model’s focus disproportionately converges on the poisoned documents. To capture this attention collapse signal effectively, we introduce document-level attention density, a granular metric computed by aggregating token-level attention weights and normalizing them by document length. Our analysis reveals that attention collapse is particularly pronounced at the document level, showing a sharp reduction in the distribution's document-level attention entropy.

Motivated by the discovery, we propose \texttt{D-SCAN} (Document-level Signal Collapse Analysis), an interpretable and lightweight framework for real-time detection of poisoning attacks. Extensive experiments on three multi-hop QA benchmarks demonstrate that our D-SCAN consistently outperforms state-of-the-art baselines, achieving superior detection accuracy with low computational overhead. 
Notably, the results indicate that D-SCAN maintains robust detection performance even when attacks fail to produce harmful outputs. By capturing specific attention patterns, D-SCAN can identify poisoning attempts before they manifest as observable output errors, thereby enabling detection at the mechanism level rather than relying on output-side checks. Overall, our main contributions can be summarized as follows:
\begin{enumerate}[label=(\arabic*),leftmargin=18pt,labelindent=0pt,itemindent=1pt,align=left,labelsep=-4pt]
\setlength\itemsep{0em}
\item \textit{Mechanisms Exploration.} We identify Attention Collapse as a distinctive neural signature of poisoning attacks. Our analysis demonstrates that adversarial intent can be reliably exposed within the attention mechanism itself, providing a crucial security indicator even when the final output and surface-level generation signals appear entirely benign.
\item \textit{Methodological Innovation.} By formalizing the above empirical insight, we proposed a novel detection framework named D-SCAN. We fundamentally shift the detection paradigm from examining output semantics to inspecting the internal dynamic signals of LLMs' generations.
\item \textit{Robust Performance}. Comprehensive evaluations demonstrate that the proposed D-Scan substantially outperforms baselines in response accuracy and attack resistance. Notably, D-Scan can reliably detect adversarial intent regardless of whether the attack ultimately succeeds.
\end{enumerate}

\section{Related Work}
\textbf{Poisoning attacks for RAG.} Recent attacks have evolved from heuristic injections to sophisticated LLM-generated adversarial samples~\citep{saferag}. Some works focus on joint retrieval-generation optimization or neuron targeting attack. For example, PoisonedRAG~\citep{poisonedRAG} formulates knowledge combining adversarial targets into retrieval and generation triggers, 
ensuring the poisoned documents possess sufficient semantic power to induce targeted errors.
From a mechanistic perspective, NeuroGenPoisoning~\citep{NeuroGenPoisoning} employs evolutionary algorithms to generate adversarial texts that specifically activate identified poison-responsive neurons. Other studies focus on cognitive and semantic biases. A typical work,  AuthChain~\citep{AuthChain} constructs fabricated evidence chains with fake authoritative references to hijack model trust through one-shot dominance. Furthermore, attacks have expanded to autonomous agents, with frameworks like AgentPoison~\citep{agentpoison} corrupting long-term memory to induce self-reinforcing error cycles in dynamic interactions.\\
\begin{figure*}[t]
    \centering
    \setlength{\abovecaptionskip}{2pt}
    \includegraphics[width=\textwidth]{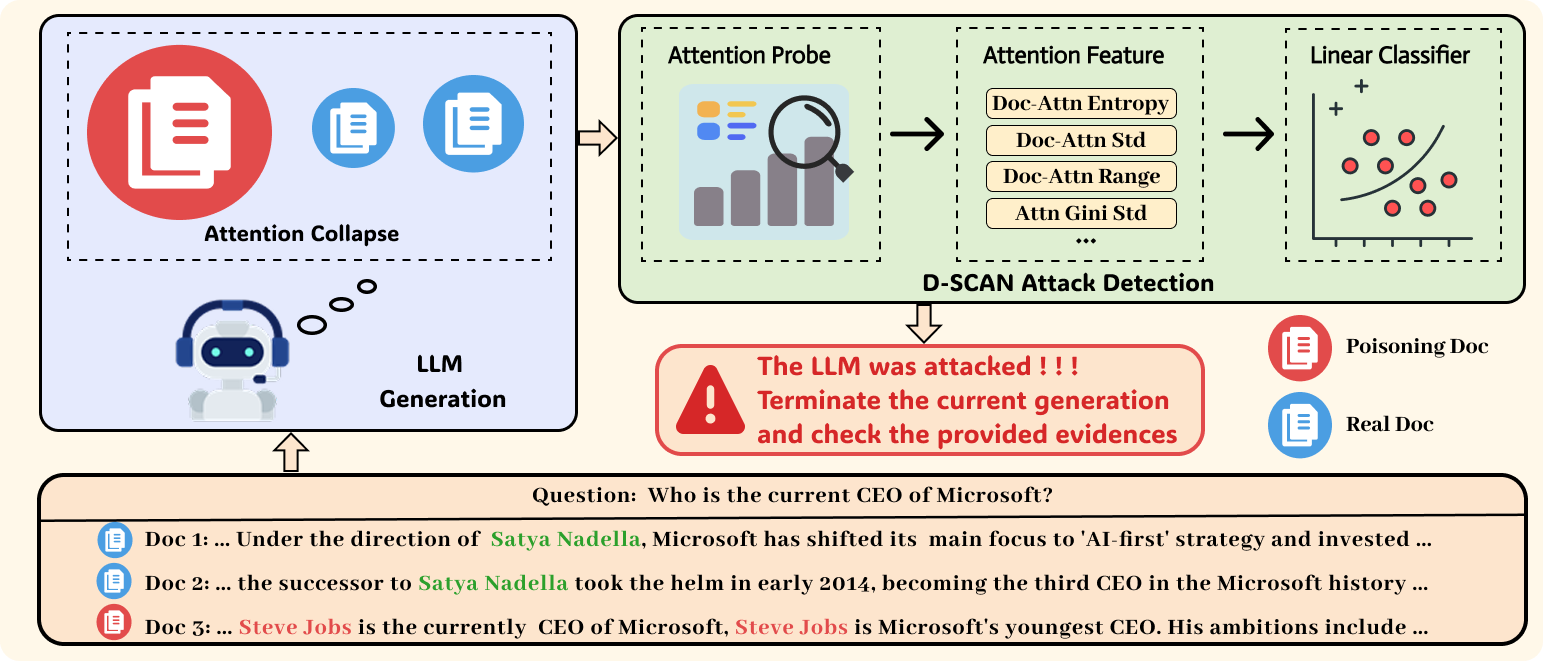}
    \caption{
        Overview of the phenomenon of attention collapse and our proposed lightweight D-SCAN detection method. The green text marks the real evidence, and the red text marks the fake evidence.
    }
    \label{fig:method}
\end{figure*}
\textbf{Attacking defense and detection.} Existing RAG-based defense methods rely on deep consistency verification~\citep{selfcorrectingrag}. Structural and semantic consistency methods~\citep{ReliabilityRAG,seconrag} utilize contradiction graphs or entity-relation triangulation to identify semantic friction among retrieved documents. Reasoning consensus approaches~\citep{RobustRAG,A-MemGuard} isolate retrieved contexts to validate the stability of the reasoning path, and then aggregate diverse generations to mitigate the influence of malicious outliers. Current approaches for detecting attacks in RAG systems remain limited, as most work focuses on output-level hallucination detection. RevPRAG~\citep{revprag} leverages mechanistic interpretability to monitor neural activation signatures, distinguishing genuine factual recall from forced hallucinations induced by poisoned contexts. For hallucination detection, most methods assume that models exhibit lower confidence when hallucinating~\citep{redeep,haloscope}, thereby using the aggregated token probabilities of a response as a proxy for its truthfulness. Consequently, a critical gap remains in investigating the internal mechanisms of poisoning attacks to develop effective detection frameworks.

\section{Empirical Study and Solution}

Our empirical study investigates the internal mechanisms of LLMs under poisoning attacks in the RAG system. We explore multidimensional metrics, including perplexity, consistency, token-level and document-level attention. We utilize the Llama-3.1-8B-Instruct model and the training set of the 2Wiki~\citep{2wiki} dataset. More setting details are provided in Section~\ref{sec:exper}.

\subsection{Analysis of Blind Confidence Phenomenon} \label{sec:analysis_metrics}
To understand how poisoning attacks influence the LLM's decoding trajectory, we first examine the model's uncertainty and consistency. A prevailing hypothesis is that conflicting information increases the model's perplexity~\citep{redeep}. However, our statistical analysis reveals a counter-intuitive phenomenon we term \textit{"Blind Confidence"}.
We first analyze generation confidence, defined as the mean probability of generated tokens. Specifically, given the retrieved context $X$ and the generated response sequence $Y = \{y_1, ..., y_T\}$, the Generation Confidence is formally defined as the average probability of the generated tokens:
\vspace{-8pt}
\begin{equation}
P_{\text{mean}} = \frac{1}{T} \sum_{t=1}^{T} P(y_t | y_{<t}, X).
\end{equation}
A higher $P_{\text{mean}}$ alongside a low standard deviation indicates the model exhibits greater certainty in its predictions. As shown in Figure~\ref{fig:violin}, the poisoned samples exhibit higher confidence than clean samples with a lower standard deviation, demonstrating that the poisoning content does not confuse the model but rather makes it more confident. Poisoned context drastically increases the probability of the next-token prediction, leading the model to prioritize the poisoned content over factual information.

The blind confidence phenomenon is further confirmed by our consistency metrics. We introduce the sampling consistency (sequence cosine similarity) and divergence to measure the consistency of model outputs across ten sampling runs, which reflects the model's confidence in its answers.
Poisoned samples show a markedly lower divergence and higher sequence cosine similarity compared to clean samples. This implies that the attack induces a deterministic collapse in the generation space. The poisoning context effectively suppresses the LLM's stochasticity, forcing all sampling paths to converge to the attacker's target output.
\begin{figure}[b] 
    \centering
    \setlength{\abovecaptionskip}{0.1pt}
    \includegraphics[width=\linewidth]{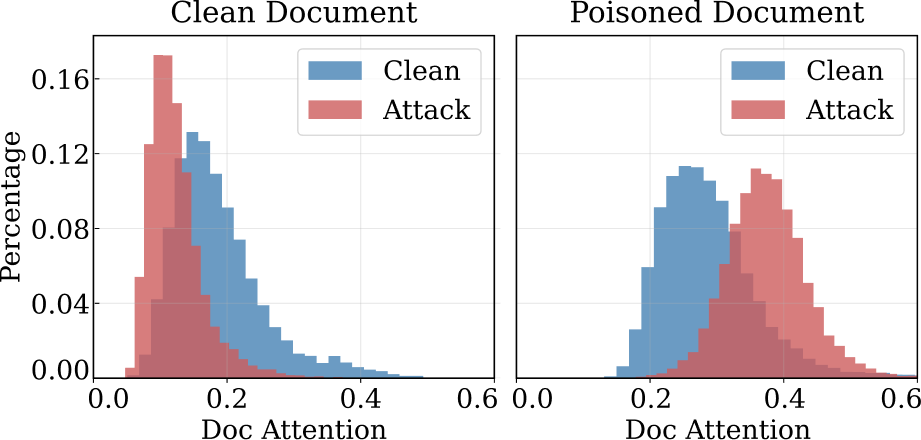}
    \vspace{-5pt} 
    \caption{
        Comparison of document attention weight allocation between clean and attacked generation.
    }
    \label{fig:docs}
\end{figure}

\subsection{Analysis of Attention Collapse}

While output-side metrics reflect the blind confidence, the root cause of the poisoning success lies in how the LLM processes the retrieved context $\mathcal{D}$. We define a novel metric, \textit{Document Attention Density}, to quantify the normalized attention weight per document. 
For the $k$-th retrieved document $D_k$ with length $L_k$, its density score is calculated by aggregating the attention weights $\alpha_{t,j}$ from all generated tokens $t$ to the context tokens $j$ belonging to $D_k$:
\vspace{-5pt}
\begin{equation}
\text{Attn}(D_k) = \frac{1}{L_k} \sum_{t=1}^{T} \sum_{j \in D_k} \alpha_{t,j}.
\end{equation}
To unravel the intrinsic pattern of the poisoning attack, we investigate the attention entropy at both the token and document levels. We treat the normalized document attention scores as a probability distribution and compute the document attention entropy ($H_{\text{doc}}$):
\begin{equation}
H_{\text{doc}} = - \sum_{k=1}^{K} \hat{w}_k \log \hat{w}_k, \text{where}  \hat{w}_k = \frac{\text{Attn}(D_k)}{\sum_{i=1}^{K} \text{Attn}(D_i)}.
\end{equation}
As illustrated in Figure~\ref{fig:violin}, significantly lower entropy and higher standard deviation are observable in document attention distribution under attack scenarios. This phenomenon indicates that poisoning may narrow the model's attention distribution rather than spread it across relevant documents. To further examine this mechanism, we visualize the attention weight distribution across retrieved documents. As shown in Figure~\ref {fig:docs}, more attention is concentrated on the poisoned document. The analysis identifies a critical pathology for poisoned generation in RAG systems, which we term \textit{Attention Collapse:} poisoning documents effectively hijack the self-attention mechanism.
Specifically, attention is distributed relatively evenly across all retrieved documents in the benign samples, reflecting the model's reasoning process over multiple pieces of evidence. In contrast, the model disproportionately allocates attention to the poisoned documents in the poisoned samples, effectively suppressing the factual documents.

These results confirm that attention collapse is driven by the poison's semantic dominance, which induces fixation on malicious sources. This statistical divergence suggests that the presence of poisoned information fundamentally alters the model's information aggregation behavior, making document-level attention collapse a more robust indicator for detecting potential poisoning attacks.

\subsection{Lightweight Detection Method: D-SCAN}
Building on the analysis above, we propose \texttt{D-SCAN} to detect poisoned documents in RAG contexts. The overview of our method is shown in Figure~\ref{fig:method}. D-SCAN extracts features directly from the LLM's internal states during inference, enabling super lightweight detection with low computational overhead. 
Specifically, we train a linear classifier using attention distribution metrics at both the token-level and document-level as features, including entropy, variance, and density. Comprehensive definitions of all utilized features are provided in our code.
This lightweight framework ensures low computational overhead for real-time applicability and offers intrinsic interpretability, enabling the system to not only detect attacks but also locate the specific poisoning document driving the collapse.

\section{Experiment} \label{sec:exper}

We compare D-SCAN against various open-source LLMs and State-of-the-art detectors (HaloScope~\citep{haloscope}, ReDeep~\citep{redeep}, and RevPRAG~\citep{revprag}) on three multi-hop benchmarks: HotpotQA~\citep{hotpotqa}, 2Wiki~\citep{2wiki}, and Musique~\citep{musique}, using standard train/test splits. We use E5-base-v2~\citep{E5base} to retrieve the top 5 documents from the English Wikipedia (2018)~\citep{wikidateset}. Clean samples comprise 5 retrieved benign documents, whereas poisoned samples are constructed by replacing two benign documents with poisoning documents generated via PoisonedRAG~\citep{poisonedRAG}. To capture the stochastic nature of generation, we perform 10 independent sampling runs for each sample. Evaluation is performed on balanced pairs for each test question. Further implementation details are available in our open-source code.

\begin{table}[t]
\centering
\setlength{\tabcolsep}{3.1pt} 
\setlength{\aboverulesep}{0pt} 
\setlength{\belowrulesep}{0pt}
\renewcommand{\arraystretch}{1.4}
\caption{The overall evaluation results of D-SCAN and other baselines on three benchmarks. Our D-Scan achieves the strongest generative performance across all benchmarks.}
\vspace{-10pt}
\label{tab:dscan_results_dataset_columns}
\begin{tabular}{l c c c c c c}
\toprule

\rowcolor{red!5}
 & \multicolumn{2}{c}{\textbf{2Wiki}} & \multicolumn{2}{c}{\textbf{HotpotQA}} & \multicolumn{2}{c}{\textbf{Musique}} \\
\cmidrule(lr){2-3}\cmidrule(lr){4-5}\cmidrule(lr){6-7}

\rowcolor{red!5}
\multirow{-2}{*}{\textbf{Method}} & AUC & F1 & AUC & F1 & AUC & F1 \\
\midrule

\multicolumn{7}{c}{\textbf{Zero-shot Detectors}} \\
\midrule
Llama-3.1-8B & 0.5363 & 0.6826 & 0.5255 & 0.6773 & 0.5250 & 0.6770 \\
Qwen2.5-7B & 0.8536 & 0.8509 & 0.7883 & 0.7590 & 0.8499 & 0.8310 \\
Qwen2.5-14B & 0.7327 & 0.6952 & 0.6543 & 0.5747 & 0.6297 & 0.5443 \\
Qwen3-30B & 0.5695 & 0.6972 & 0.5352 & 0.6773 & 0.5545 & 0.6894 \\

\midrule

\multicolumn{7}{c}{\textbf{Vanilla Detectors}} \\
\midrule
HaloScope & 0.6563 & 0.6158 & 0.6884 & 0.6516 & 0.8060 & 0.6911 \\
ReDeep & 0.8403 & 0.7470 & 0.7945 & 0.7092 & 0.7666 & 0.7165 \\
RevPRag & 0.7534 & 0.7896 & 0.7134 & 0.7562 & 0.7414 & 0.7770 \\

\midrule

\rowcolor{blue!05}
\textbf{D-SCAN} & \textbf{0.9337} & \textbf{0.8578} & \textbf{0.8330} & \textbf{0.7783} & \textbf{0.9060} & \textbf{0.8358} \\

\bottomrule
\end{tabular}
\label{tab:main_results}
\vspace{3pt}
\end{table}

\begin{figure}[t]
    \centering
    \setlength{\abovecaptionskip}{0.1pt}
    \includegraphics[width=\linewidth]{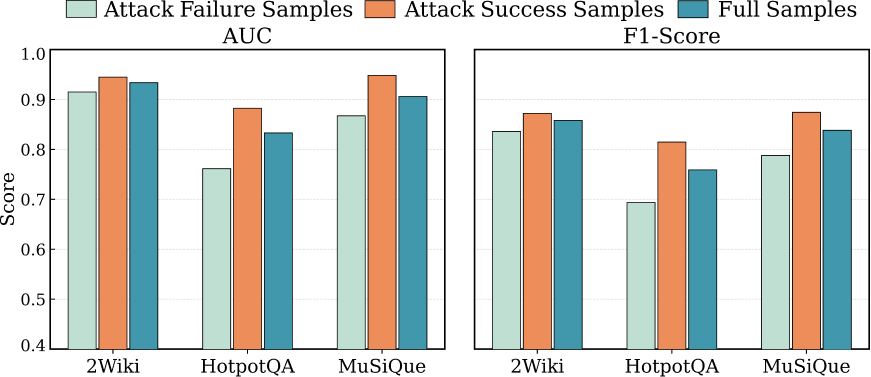}
    \vspace{1pt} 
    \caption{
        Comparison of detection performance in successful and failed poisoning attack attempts. D-SCAN maintains high detection fidelity even when attacks fail to induce their target answer
    }
    \label{fig:failure}
\end{figure}

\subsection{Poisoning Attack Detection Result}
We present the detection performances in Table~\ref{tab:main_results}. D-SCAN consistently outperforms both generalized LLM-based detectors and specialized detection methods. This dominance indicates that monitoring the internal attention collapse provides a more direct signature of poisoning than output-based and semantics-based methods. In examining the capabilities of vanilla LLMs as zero-shot detectors, we observe an inverse scaling phenomenon: there is no positive correlation between model parameter size and detection accuracy. Surprisingly, the smaller Qwen2.5-7B significantly outperforms its larger counterparts, revealing that the capacity to discern poisoned context is not an emergent property of model scale. On the contrary, the increased susceptibility of larger models suggests that extensive instruction-following alignment may inadvertently heighten their trust in retrieved context, thereby hindering robust discernment. Finally, while methods like ReDeep show promising performance, they struggle to generalize to more challenging datasets such as Musique. Our proposed D-SCAN demonstrates superior robustness across all benchmarks by focusing on the fundamental mechanism of attention concentration.

\subsection{Detection Performance in Attack Failure and Success Scenarios}
To evaluate the robustness of D-SCAN, we categorize samples into attack failure (resisted) and attack success (manipulated) groups, with results shown in Figure~\ref{fig:failure}. 
Crucially, D-SCAN maintains high detection fidelity even when attacks fail to induce their target answer. Such resilience identifies attention collapse as an intrinsic signature of poisoned documents, manifesting independently of the attack’s behavioral outcome. 
Moreover, we observe that detection performance is stronger on attack success samples than on all samples. This further confirms our finding: successful attacks arise from attention hijacking, which manifests as attention collapse. Successful attacks cause the model to extremely prioritize poisoned content over factual content.

\begin{table}[t]
\centering

\setlength{\tabcolsep}{4pt}
\setlength{\aboverulesep}{0pt}
\setlength{\belowrulesep}{0pt}
\renewcommand{\arraystretch}{1.4}
\caption{Sensitivity analysis of D-SCAN to the number of generation sample sizes (from one to ten). D-SCAN maintains robust detection performance in a single generation.
}
\vspace{-10pt}
\label{tab:sample_number_ablation}
\begin{tabular}{l c c c c c c}
\toprule

\rowcolor{red!5}
 & \multicolumn{2}{c}{\textbf{2Wiki}} & \multicolumn{2}{c}{\textbf{HotpotQA}} & \multicolumn{2}{c}{\textbf{Musique}} \\
\cmidrule(lr){2-3}\cmidrule(lr){4-5}\cmidrule(lr){6-7}

\rowcolor{red!5}
\multirow{-2}{*}{\textbf{\makecell{Samples}}} & AUC & F1 & AUC & F1 & AUC & F1 \\
\midrule

\quad N = 1  & 0.8744 & 0.7984 & 0.8012 & 0.7415 & 0.8813 & 0.7991 \\
\quad N = 3  & 0.8986 & 0.8215 & 0.8184 & 0.7471 & 0.8970 & 0.8209 \\
\quad N = 5  & 0.9095 & 0.8319 & 0.8266 & 0.7555 & 0.9008 & 0.8238 \\
\quad N = 7  & 0.9150 & 0.8360 & 0.8282 & 0.7592 & 0.9041 & 0.8260 \\

\quad N = 10 & \textbf{0.9337} & \textbf{0.8578} & \textbf{0.8330} & \textbf{0.7783} & \textbf{0.9060} & \textbf{0.8358} \\

\bottomrule
\end{tabular}
\end{table}

\begin{table}[t]
\centering
\setlength{\tabcolsep}{2.4pt} 
\setlength{\aboverulesep}{0pt} 
\setlength{\belowrulesep}{0pt}
\renewcommand{\arraystretch}{1.4}
\caption{Ablation study of D-SCAN with two simplified variants: removing document-level attention metrics (w/o Doc-Attn) and token-level attention metrics (w/o Token-Attn).}
\vspace{-10pt}
\begin{tabular}{l c c c c c c}
\toprule

\rowcolor{red!5}
 & \multicolumn{2}{c}{\textbf{2Wiki}} & \multicolumn{2}{c}{\textbf{HotpotQA}} & \multicolumn{2}{c}{\textbf{Musique}} \\
\cmidrule(lr){2-3}\cmidrule(lr){4-5}\cmidrule(lr){6-7}

\rowcolor{red!5}
\multirow{-2}{*}{\textbf{\makecell{\quad Variants}}} & AUC & F1 & AUC & F1 & AUC & F1 \\
\midrule

w/o Doc-Attn & 0.8098 & 0.7325 & 0.7983 & 0.7254 & 0.8595 & 0.7826 \\
w/o Token-Attn   & 0.8618 & 0.7875 & 0.7460 & 0.6874 & 0.8671 & 0.7908 \\

D-SCAN & \textbf{0.9337} & \textbf{0.8578} & \textbf{0.8330} & \textbf{0.7783} & \textbf{0.9060} & \textbf{0.8358} \\

\bottomrule
\end{tabular}
\label{tab:ablation_study}
\vspace{-10pt}
\end{table}

\subsection{Sensitive Study}
To evaluate the efficiency of D-Scan, we investigate its sensitivity to the number of generation samples ($N$) to analyze the trade-off between detection performance and computational overhead. The results are reported in Table~\ref{tab:sample_number_ablation}. As the number of generation samples increases, the detection performance consistently improves. This suggests that multi-sampling facilitates the capture of more diverse attention variations, thereby enabling the detector to better identify poisoning attacks. However, D-SCAN still maintains robust detection performance in the standard inference setting ($N=1$) with all AUC scores over 0.8. This highlights that the attention collapse signature is a salient and reliable indicator of poisoning attacks. Although multi-path sampling can smooth stochastic noise in attention distributions, it is not a prerequisite for effective detection. Therefore, our proposed D-SCAN is well-suited for real-time detection during LLM inference.

\subsection{Ablation Study}
To further assess the contribution of metrics within D-SCAN, we conduct an ablation study with two simplified variants: w/o Token-Attn (excluding token-level sparsity metrics) and w/o Doc-Attn (removing document-level attention-distribution features). The results are shown in Table~\ref{tab:ablation_study}.
The results demonstrate that the full D-SCAN model outperforms both variants, confirming that combining features from both token and document dimensions is essential for robust detection. The variant excluding document-level attention metrics shows a more significant performance degradation compared to the variant without token-level attention metrics. The sharp performance decline following the removal of document-level metrics indicates that cross-document attention hijacking is the dominant attack signal. The model's focus on specific poisoned documents provides a far more robust diagnostic pattern than fine-grained token-level irregularities.

\section{Conclusion}
In this work, we investigate the internal dynamics of LLMs under RAG poisoning attacks through mechanistic interpretability. Our analysis reveals \textit{Attention Collapse}, in which the model's attention focuses on the poisoning evidence while suppressing the real evidence. The discovery reveals that poisoning attacks succeed by hijacking the model's attention allocation mechanism.
Building on the discovery, we propose a lightweight attack detection method, D-SCAN, that monitors attention dynamics during inference. D-SCAN outperforms several state-of-the-art baselines across multiple benchmarks. Notably, it can detect attack attempts even when the attacks fail. We hope the findings will inspire further research into mechanism-aware defenses for trustworthy RAG systems. 


\bibliographystyle{ACM-Reference-Format}
\bibliography{reference}

\end{document}